\documentclass[aps,pre,twocolumn,showpacs,superscriptaddress,eqsecnum]{revtex4}
\usepackage{amsmath,graphicx,amssymb}

\begin{document}

\title{Growth of surface undulations at the Rosensweig instability}
\author{Holger Knieling}
\affiliation{Experimentalphysik V, Universit\"at Bayreuth, D-95440 Bayreuth, Germany}
\author{Reinhard Richter}
\affiliation{Experimentalphysik V, Universit\"at Bayreuth, D-95440 Bayreuth, Germany}
\author{Ingo Rehberg}
\affiliation{Experimentalphysik V, Universit\"at Bayreuth, D-95440 Bayreuth, Germany}
\author{Gunar Matthies}
\affiliation{Ruhr-Universit\"at Bochum, Universit\"atsstra{\ss}e 150, D-44780 Bochum, Germany}
\author{Adrian Lange}
\affiliation{Fraunhofer Institute for Material and Beam Technology, Winterbergstra{\ss}e 28, D-01277 Dresden, Germany}

\date{Received 14 May 2007; published 4 Dezember 2007}

\begin{abstract}
We investigate the growth of a pattern of liquid crests emerging in
a layer of magnetic liquid when subjected to a magnetic field
oriented normally to the fluid surface. After a steplike increase
of the magnetic field, the temporal evolution of the pattern
amplitude is measured by means of a Hall-sensor array. The
extracted growth rate is compared with predictions from linear
stability analysis by taking into account the proper nonlinear
magnetization curve $M(H)$. The remaining discrepancy can be
resolved by numerical calculations via the finite-element method.
By starting with a finite surface perturbation, it can reproduce the
temporal evolution of the pattern amplitude and the growth rate.
The investigations are performed for two magnetic liquids, one with
low and one with high viscosity.
\end{abstract}

\maketitle

\section{\label{sec:intro}Introduction}

Plato (c. 427-347 B.C.) remarked: ``You know that the beginning is the
most important part of any work, especially in the case of a young and
tender thing; for that is the time at which the character is being
framed'' \cite{plato}. The same may be true in pattern formation,
which makes it most rewarding to look at its early stage. At the beginning
of an evolving pattern stands an unstable mode \cite{cross93}. As long as
the amplitude of the mode is small, its wave number and growth rate can
be calculated by linear stability analysis. In this way the early stage
of pattern formation has been investigated in many different systems.

Considering interface instabilities, the Rayleigh-Taylor
configuration is the most prominent example. Here the growth rate of
the fastest-growing mode has been measured for granular suspensions
\cite{voeltz01} and for immiscible fluids \cite{carles06}. In the
latter case a monotonic, roughly linear dependence of the growth
rate as a function of the density difference was derived and
observed. The difficulty in setting experiments with the
Rayleigh-Taylor instability is that the driving gravitational field
can not be switched on externally. This makes the preparation of a
plane layer as a starting condition cumbersome.

This difficulty is eluded if the interface instability is driven by
an externally applied electric or magnetic field. For an
electrohydrodynamic instability of a polymer liquid/air interface the
growth rate of the dominant mode was recently measured to increase
with the sixth power of the reduced electrical field \cite{leach05},
as predicted by linear stability analysis for thin films. These thin
films show a monotonic dispersion relation. However, the situation
is different for thick layers, where the weight of the liquid has to
be taken into account. This results in gravitational waves, leading
to a nonmonotonic dispersion relation \cite{taylor65,cowley67}.
Neither for the electrostatic interface instability (see, e.g.,
$^3$He-$^4$He mixtures \cite{wanner79}) nor for its magnetostatic
counterpart, has the growth rate of the linearly most unstable mode been
measured hitherto. In the following we fill this gap for the
magnetostatic case.

\begin{figure}[tbp]
\begin{center}
  \includegraphics[width=0.99\columnwidth]{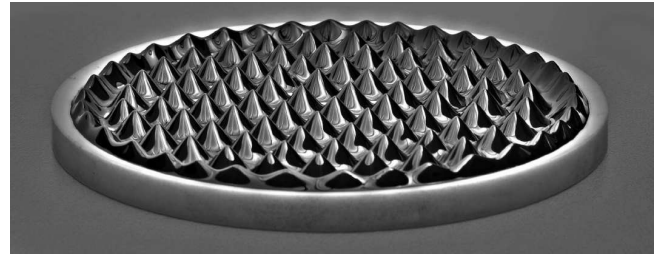}
  \caption[]{Rosensweig peaks of the magnetic fluid type EMG 909,
  Ferrotec Co., at a supercritical induction $B>B_c$ in a vessel
  with diameter of 120\,mm. A movie showing the formation of
  Rosensweig patterns can be accessed at \protect
  Ref.\,\cite{castellvecchi05}.}
  \label{fig:peaks}
\end{center}
\end{figure}

The Rosensweig or normal field instability \cite{cowley67} is
observed in a layer of magnetic fluid (MF) \cite{rosensweig85}, when
a critical value $B_c$ of the vertical magnetic induction is
surpassed. Figure \ref{fig:peaks} presents a photo of the final
pattern of static liquid peaks, which emerge due to a transcritical
bifurcation. This was investigated in theory
\cite{gailitis77,friedrichs01,friedrichs03} and experiments
\cite{bacri84,richter05,gollwitzer07}. For a sudden increase of the
magnetic induction $B$ the \textit{wave number} $q_m$ of the
fastest-growing mode was measured in the linear range, i.e. for small
amplitudes \cite{abou00,lange00_wave,reimann03,lange07}. In
agreement with theory its value increases monotonically with the
supercritical magnetic induction. The growth rate of the
fastest-growing mode was recently calculated in detail
\cite{lange01_growth}. Here we present an experimental test
of those predictions.

In order to measure the temporal evolution of the growing amplitudes
we utilize a linear array of Hall sensors \cite{reimann05}, which is
sketched together with the experimental arrangements in
Sec.\,\ref{sec:experiment}. The results are compared with the
outcome of the linear stability analysis in Sec.\,\ref{sec:theory}
and with numerical calculations in Sec.\,\ref{sec:numerics}.

\section{\label{sec:experiment}Experiment}

\begin{figure}[htbp]
  \begin{center}
  (a)\includegraphics[width=0.45\textwidth]{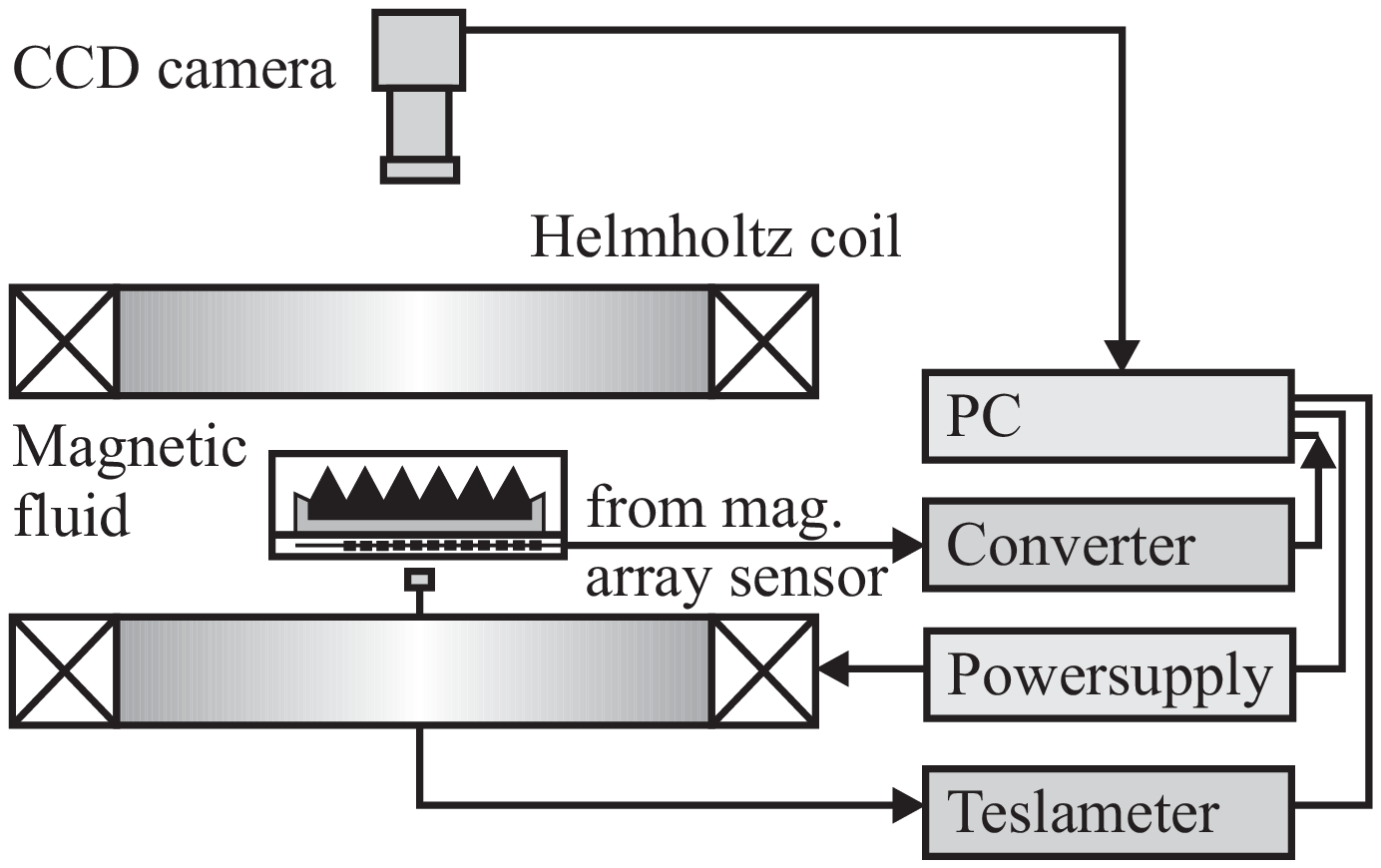}
  ~\vskip 0.3cm
  (b)~~~~~~~\includegraphics[width=0.36\textwidth]{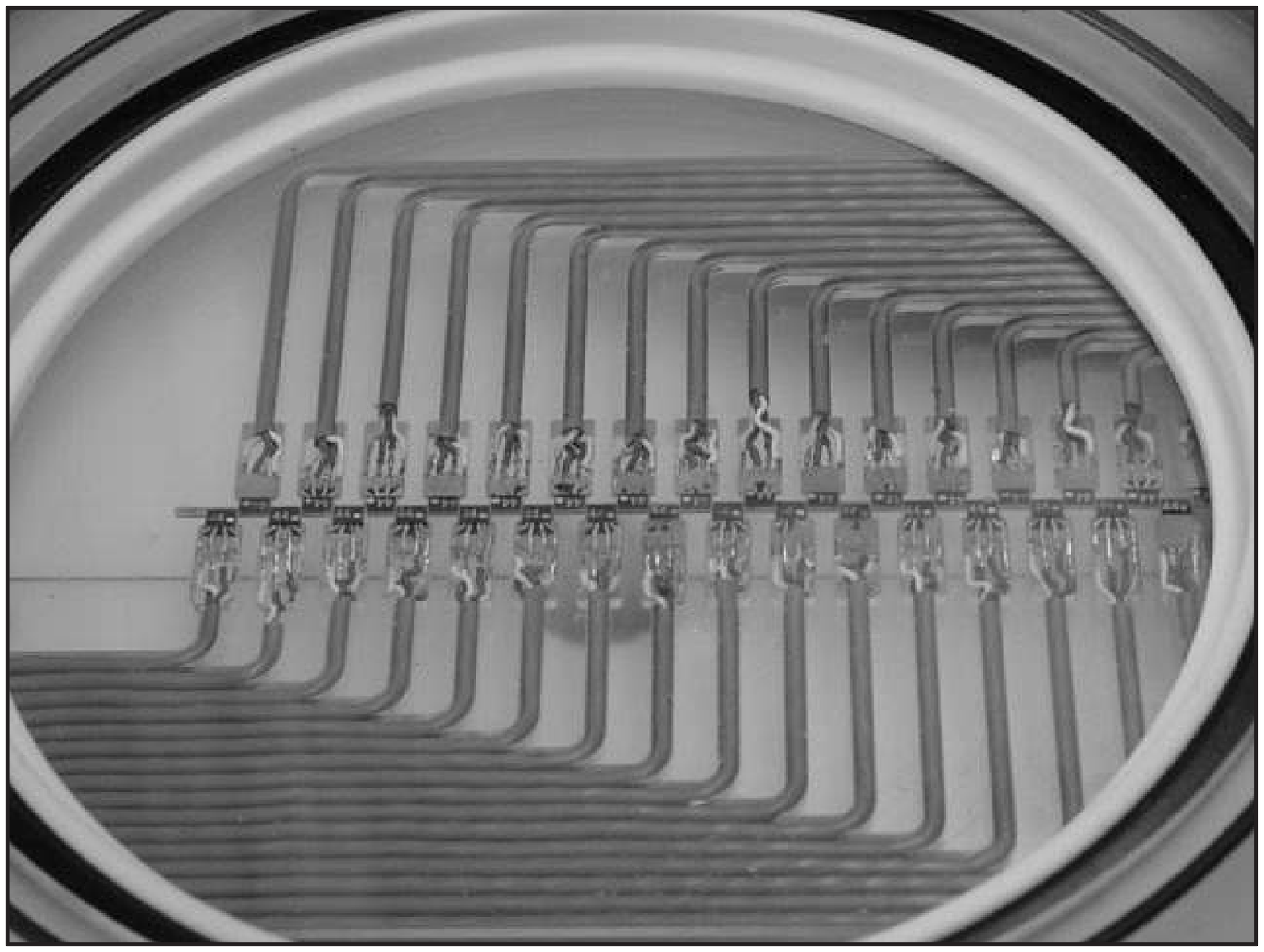}
  \caption{Magnetic measuring principle: (a) Sketch of the experimental
  setup; (b)~photograph of the linear array of 32 Hall sensors mounted
  1.78 mm under the bottom of a transparent vessel.}
  \label{fig:setup}
  \end{center}
\end{figure}

Our experimental setup is shown in Fig.~\ref{fig:setup}(a). A
cylindrical vessel with an edge made of Teflon with a radius of
60 mm and a depth of 5 mm is filled to the brim with the
MF and situated in the center of a Helmholtz pair of coils (for
details see Ref.\,\cite{reimann03}). A camera is positioned above
the vessel for optical observation. For calibration purposes a
commercial Hall probe (Group3-LPT-231) in combination with a
digital teslameter (DTM 141) was used. For measuring the temporal
evolution of the surface amplitude we take advantage of the local
variation of the magnetic field, which is increased immediately
beneath a magnetic spike and reduced beneath the interspike area.
In order to measure these local variations, a linear array of 32
Hall sensors (KSY 44, Siemens Co.) was mounted $1.78\pm 0.1$ mm
below the bottom of the dish, as shown in Fig.~\ref{fig:setup}(b).
The sensors communicate via 32 amplifiers and a bus with the PC.
Details of this method are presented in Ref.\,\cite{reimann03_phd}.
In this way line scans with a frequency up to 7 kHz are possible.
This time resolution makes the method suitable for measurement of
the growth rate of the pattern evolution. Although this technique
is superior to the radioscopic method \cite{richter01} in terms
of speed, and this was our main reason for selecting it for our
purposes, we should also mention its disadvantages such as the
limited vertical ($1\;\mu$T) and lateral (3.2 mm) resolution.

The experiments are performed with the magnetic fluids EMG 909 (Lot No.
F050903B) and APG J12 (Lot No. F112795C) from Ferrotec Co. Their material
parameters were measured and are as follows: a density of
$\rho = 1005\,(1097)\,\mathrm{kg\,m^{-3}}$, a surface tension of
$\sigma = 2.4\times 10^{-2}\,(2.89\times 10^{-2})\,\mathrm{N\,m^{-1}}$,
and a dynamic viscosity of $\eta = 4.2\times 10^{-3}\,(51.9\times
10^{-3})\,\mathrm{Pa\,s}$. The parameters of EMG 909 differ slightly
from those in Ref.\,\cite{reimann03} because of a new method of
fabrication of that fluid.

These two test fluids were chosen because their material parameters
are rather similar, with one exception: the dynamic viscosity differs
by nearly an order of magnitude. By carrying out the measurements
for both fluids one can judge whether the viscosity influences the
degree of agreement in a comparison between theory, numerics, and
experiment with respect to the growth rate.

\begin{figure}[htbp]
  \begin{center}
  \includegraphics[scale=0.8]{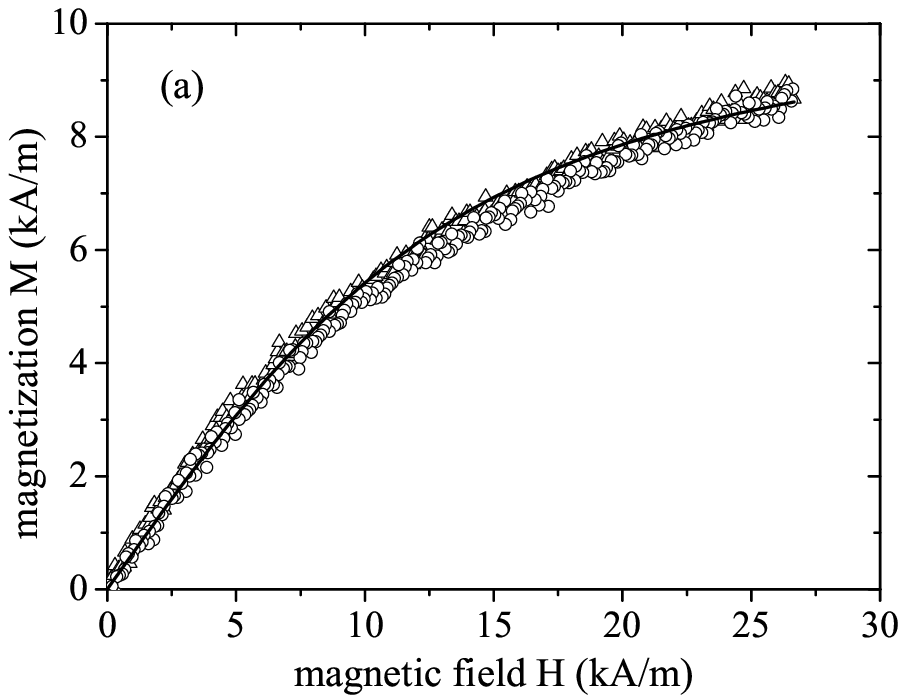}
  \includegraphics[scale=0.8]{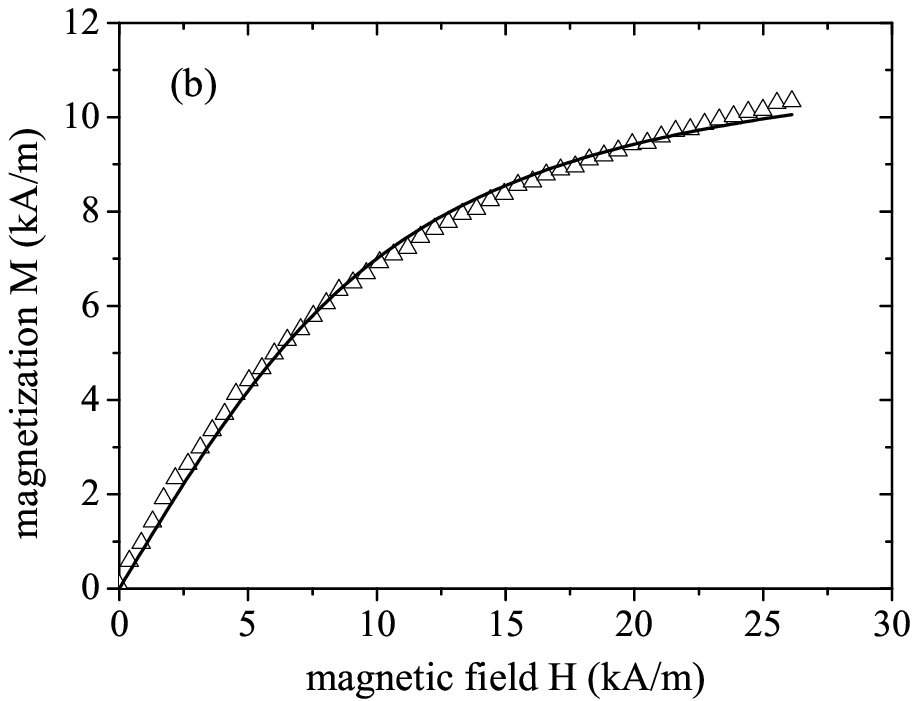}
  \caption{Magnetization $M$ versus the magnetic field $H$ for the
  magnetic fluids EMG 909 (a) and APG J12 (b). The triangles indicate
  $M$ for an increasing field, and the open circles for a
  decreasing field. The solid line gives the fit with the simple
  Langevin function (see text).}
  \label{fig:magnetization}
  \end{center}
\end{figure}

Furthermore, the magnetization curve $M=M(H)$ was measured (see
symbols in Fig.~\ref{fig:magnetization}). To exploit the experimental
data for the theoretical calculations, the points can be fitted in
the investigated range \cite{browaeys99} with a simple Langevin
function,
\begin{equation}
  L(\alpha) = M^{\displaystyle{^\star}}_s\left(\coth(\alpha)-\frac{1}{\alpha}\right)
  \text{ with }
  \alpha = \frac{3\chi_0}{M^{\displaystyle{^\star}}_s}\,H \; .
  \label{eq:langevin}
\end{equation}
The best fit for EMG 909 (APG J12) yields a saturation magnetization of
$M^\star_s = 10.92\;(12.12)\,\mathrm{kA\,m^{-1}}$ and an initial
susceptibility of $\chi_0$ = 0.65\;(0.91) (see the solid lines in
Fig.~\ref{fig:magnetization}). Here $M^\star_s$ denotes a value that
serves for a convenient description of the magnetization in the
low-field regime. $M^\star_s$ differs from the true saturation
magnetization $M_s$ obtained from the entire range of magnetic fields.
That range should be fitted with a more advanced function, which takes
into account also the polydisperse nature of the MF
(see Ref.\,\cite{richter07}, Chap.~3.8).

The data above lead to theoretical values for the critical induction
\cite{rosensweig85} of $B_{\rm c, theor}$ = 24.9 mT for EMG 909 and
$B_{\rm c, theor}$ = 20.3 mT for APG J12. The experimental values
were measured as $B_{\rm c, exp}$ = 25.7 (21.7) mT for EMG 909
(APG J12), which is a quite good agreement with a difference of
only about 3\,\% (6\,\%).

On switching on the Helmholtz coils in a jumplike manner, the magnet
will need a finite response time $\tau_B$. This time depends on the
size of the jump $\Delta B$ and has a maximum of $\tau_B = 30$ ms for a
maximal jump of $\Delta B = 35$ mT. To reduce this time, we start all
measurements from a subcritical induction of $B_\mathrm{sub} =
0.84\,B_c$, which leads to $\tau_B \approx 10\,\mathrm{ms}$. The
other characteristic times are the capillary time $t_c =
\sigma^{1/4}/(g_{0}^{3/4}\rho^{1/4}) \simeq
12.6\,(12.9)\,\mathrm{ms}$ and the viscous time, $t_{\nu} =
\sigma/(\rho g_0\nu) \simeq 583\,(57)\,\mathrm{ms}$, with the fluid
parameters as listed above for EMG 909 (APG J12). The kinematic
viscosity $\nu$ is given by $\eta/\rho$.

\begin{figure}[htbp]
\begin{center}
  \includegraphics[scale=0.85]{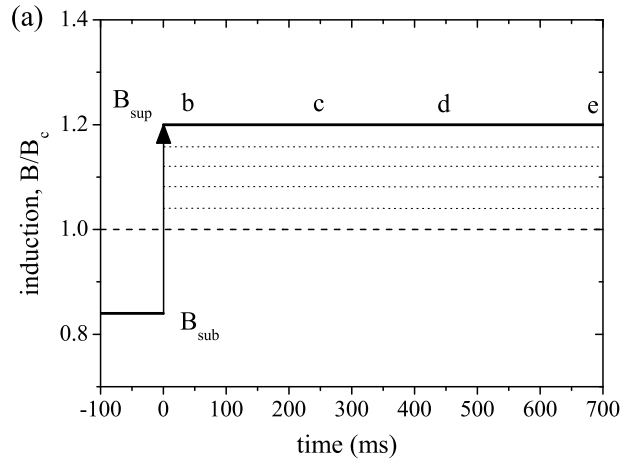}
  ~\vskip 0.2cm
  \includegraphics[width=\columnwidth]{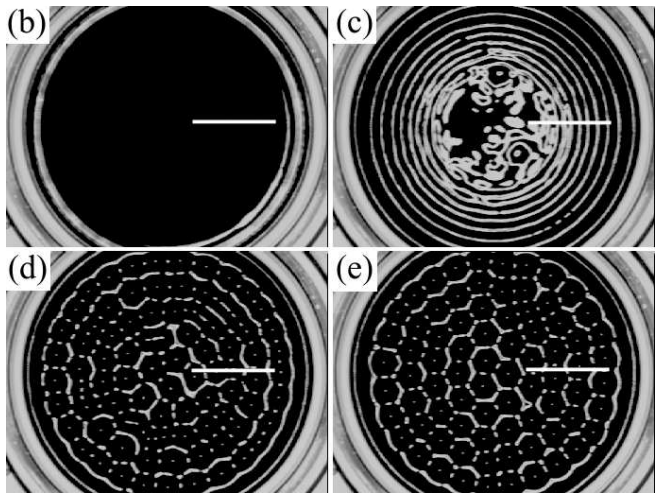}
  ~\vskip 0.0cm
  \includegraphics[width=\columnwidth]{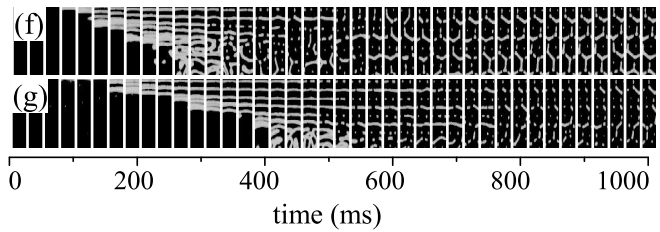}
  \caption{Measuring the growth rate of the normal field
  instability. (a) Pulse sequence. The full lines display the jump
  from a sub- to a supercritical magnetic induction. The small
  letters b, c, d, and e mark the times when the pictures (b)--(e)
  were captured. These snapshots show the pattern at times of 40 (b),
  250 (c), 450 (d), and 700 ms (e) for EMG 909. The white horizontal
  lines in the pictures indicate the area of measurement. (f), (g)
  display the pattern evolution within a small area ($8.2\times 32.7
  \rm{mm}^2$) around this location. (f) presents a sequence of images
  for the MF mark EMG 909, and (g) the corresponding sequence for the
  MF mark APG J12.}
  \label{fig:method}
\end{center}
\end{figure}

Figure\,\ref{fig:method} demonstrates the utilized magnetic pulse
sequence [Fig.\;\ref{fig:method}(a)] and the evolution of the
surface structure [Figs.\;\ref{fig:method}(b)--\ref{fig:method}(g)].
As shown in Fig.\;\ref{fig:method}(a), the magnetic induction is
jumplike increased from a sub- to a supercritical value at time
$t=0$ ms. From Fig.~\ref{fig:method}(b) we deduce that the surface
deformations first emerge at the edge of the vessel. This is due to
the discontinuity of the magnetic induction at this place. Because
of this inhomogeneous growth of the amplitude across the vessel, the
amplitude is measured only in a small region of about 35 mm between
the edge and the center of the container, as marked by white
horizontal lines in Figs.~\ref{fig:method}(b)--\ref{fig:method}(e).
Figure \ref{fig:method}(f) displays the evolution of the pattern
from a stripe-like to a hexagonal arrangement in the area of
measurement. Whereas Figs.\;\ref{fig:method}(b)--\ref{fig:method}(f)
were recorded for the fluid EMG 909, we display in
Fig.\;\ref{fig:method}(g) the pattern evolution for APG J12.
Its surface undergoes similar stages; only the time of appearance
of those structures is different.

\begin{figure}[htbp]
\begin{center}
  \includegraphics[scale=0.8]{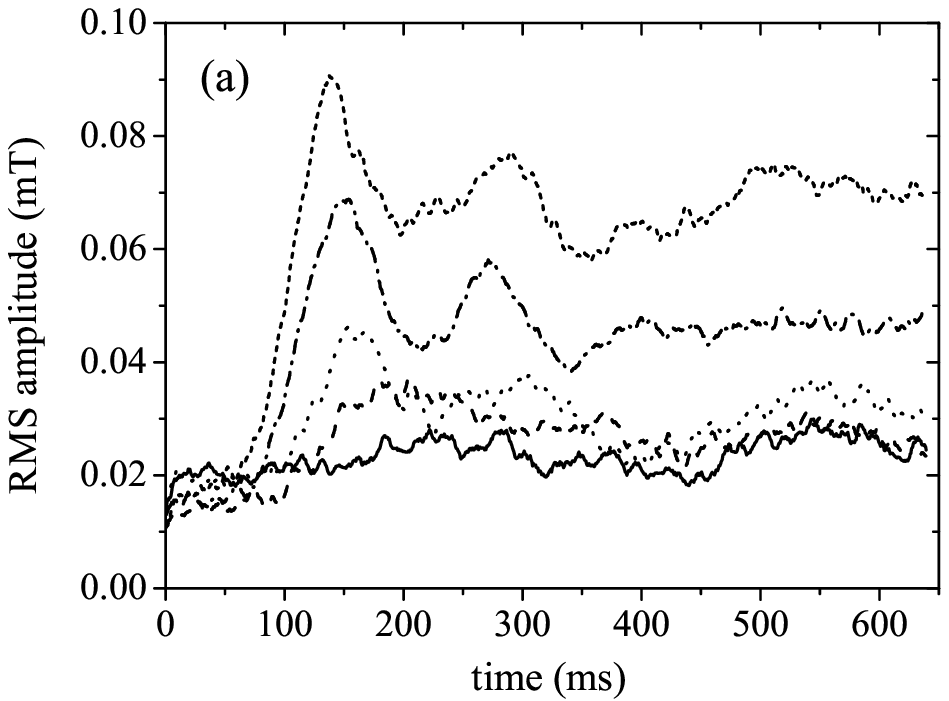}
  \phantom{AA$\,\,$}\includegraphics[scale=0.8]{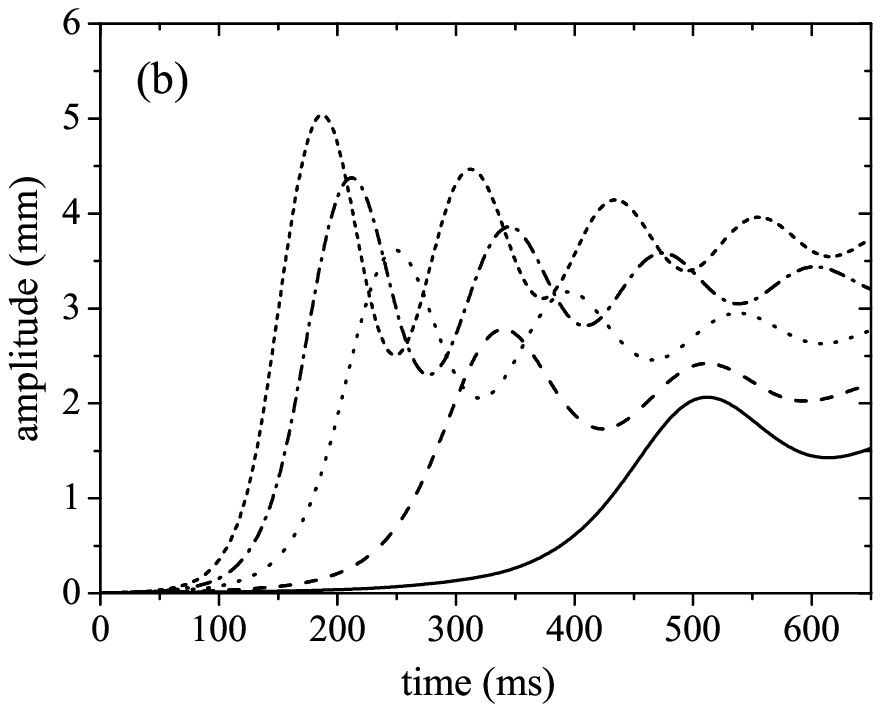}
  \caption{Time-resolved amplitudes for the fluid EMG 909. (a) Measurements
  for increasing supercritical inductions $\hat{B} = 0.028$ (full line),
  $0.057$ (dashed line), $0.100$ (dotted line), $0.157$ (dash-dotted line),
  and $0.200$ (short-dashed line). For clearer appearance, the plotted
  lines are smoothed by averaging ten neighboring points of the original
  data set. (b) Numerical results for $\hat{B} = 0.028$ (full line),
  $0.058$ (dashed line), $0.103$ (dotted line), $0.153$ (dash-dotted line),
  and $0.203$ (short-dashed line).}
  \label{fig:EMG909_amplitude}
\end{center}
\end{figure}

\begin{figure}[htbp]
\begin{center}
  \includegraphics[scale=0.8]{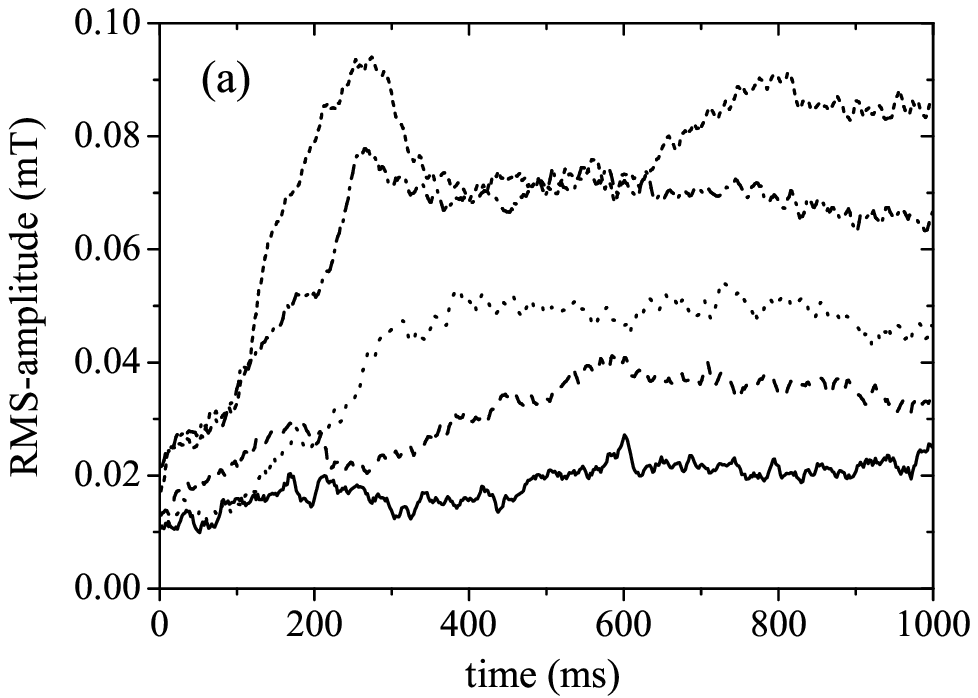}
  \phantom{AA$\,$}\includegraphics[scale=0.8]{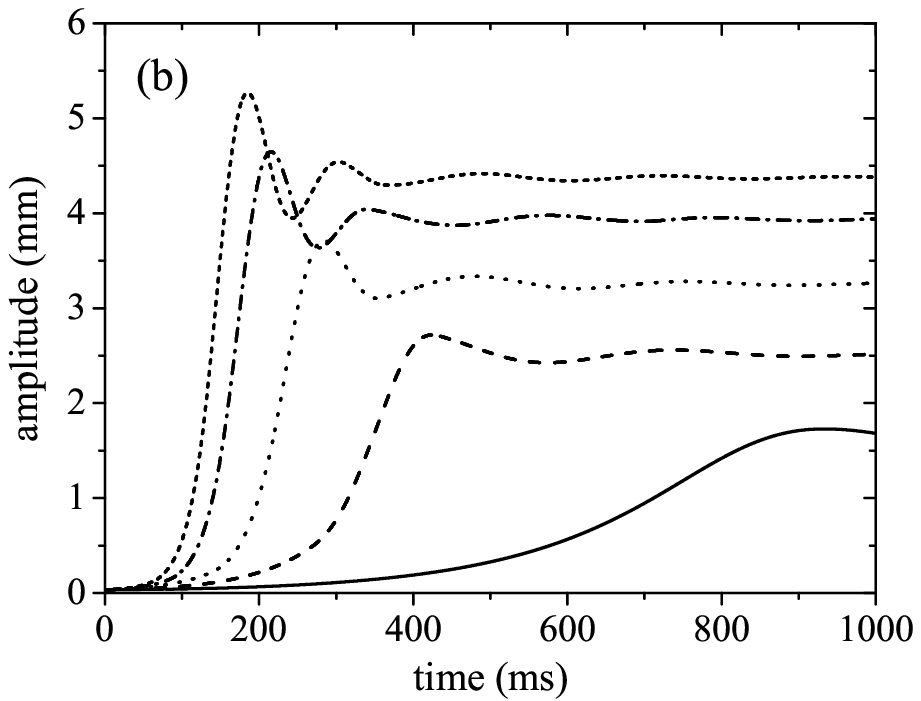}
  \caption{Time-resolved amplitudes for the fluid APG J12. (a) Measurements
  for increasing supercritical inductions $\hat{B} = 0.025$ (full line),
  $0.082$ (dashed line), $0.152$ (dotted line), $0.236$ (dash-dotted line),
  and $0.300$ (short-dashed line). For clearer appearance, the plotted
  lines are smoothed by averaging five neighboring points of the original
  data set. (b) Numerical results for $\hat{B} = 0.024$ (full line),
  $0.082$ (dashed line), $0.151$ (dotted line), $0.237$ (dash-dotted line),
  and $0.305$ (short-dashed line).}
  \label{fig:APGJ12_amplitude}
\end{center}
\end{figure}

On the basis of the time-resolved measured data points of the sensor
array, we determine the amplitude from the root-mean-square value
(rms) of that data. We display the result for EMG 909 in
Fig.\,\ref{fig:EMG909_amplitude}(a) and for APG J12 in
Fig.~\ref{fig:APGJ12_amplitude}(a). For these measurements the
induction was increased from the subcritical value $\hat{B} =
(B-B_c)/B_c = -0.16$ to supercritical values in the interval from
$\hat{B}$ = 0.0 to 0.3. The offset of the amplitude results from the
noise of the Hall sensors. The first phase of growth shows a
dramatic increase, which is followed by an oscillatory relaxation
towards the final stage in the pattern-forming process. That
relaxation process differs from a purely damped sinusoidal one due
to the reorganization of the peaks into a hexagonal pattern.

The corresponding outcome of the numerical simulations (see Sec.
\ref{sec:numerics}) is presented in
Fig.~\ref{fig:EMG909_amplitude}(b) for EMG 909 and
Fig.~\ref{fig:APGJ12_amplitude}(b) for APG J12. These plots show the
height of the amplitude with time, as calculated before
\cite{matthies05}, but for the parameters of the
investigated MF. A drastic increase of the surface height is followed
by an oscillatory relaxation, in remarkable agreement with the
measurements. The less viscous fluid EMG 909 goes through several
oscillations after a steep increase, whereas the more viscous fluid
APG J12 goes through very few oscillations.

Next we describe the extraction of the growth rate from the amplitude
curves in Figs.~\ref{fig:EMG909_amplitude}(a)
and~\ref{fig:APGJ12_amplitude}(a). The first phase of growth in the
amplitude is fitted with $y(t) = y_0 + A\exp (\omega_2\,t)$, where
$y_0$ denotes an offset and $A$ the amplitude of the exponential growth.

\begin{figure}[htbp]
\begin{center}
  \includegraphics[scale=0.8]{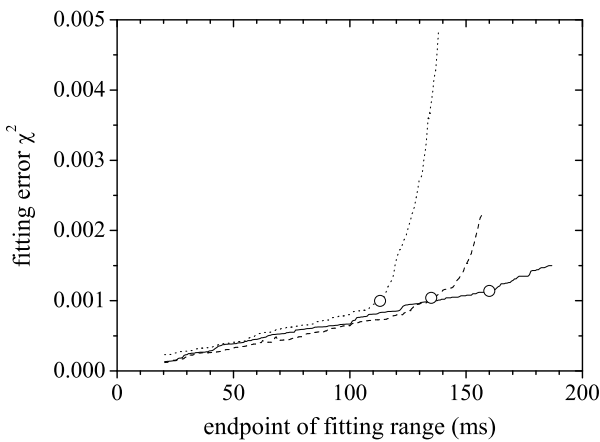}
  \caption{Errors for fits of three amplitude curves with
  magnetic inductions of $\hat{B}$ = 0.05 (full line), 0.1 (dashed
  line), and 0.2 (dotted line) in dependence on the endpoint of the
  fitting range. The open circles mark the end of the fitting range.}
  \label{fig:error}
\end{center}
\end{figure}

Due to the noisy experimental data it is difficult to determine the
area of validity for the exponential growth. Therefore we adopt the
following procedure. First we fit the offset $y_0$ of the amplitude
in the range $t = [0, 20]$ ms for EMG 909 ($t =[0, 50]$ ms for APG
J12) with a straight line without slope and hold this value constant
in the following fits. Next, a series of fits of the amplitude curve
with an exponential function is performed, where the endpoint of
the fitting range is varied in the interval from $t$ = 20 ms to the
time when the amplitude reaches its maximum. We estimate the end of
the exponential range from the evolution of the fitting error
$\chi^2$ according to Fig.~\ref{fig:error}. This value increases
linearly as more data points are considered as long as the fitted
curve is well described by an exponential function. The maximal
fitting range is reached when $\chi^2$ deviates from the linear
increase and grows with a much higher rate than before. The
beginning of this deviation indicates the proper fitting range for
the maximal growth rate, as marked in Fig.~\ref{fig:error} by open
circles for three curves at magnetic inductions of $\hat{B}$ = 0.05,
0.1, and 0.2. With increasing induction the deviation from the
linear growth of $\chi^2$ becomes more prominent. The end of the
fitting range estimated in this way is in accordance with the
inflection point determined by visual inspection from the temporal
evolution of the amplitude. The error in the growth rate resulting
from the uncertainty of the fitting range was tested to be about
10\,\% of the value of the growth rate for all applied inductions.

\begin{figure}[tbph]
\begin{center}
  \includegraphics[scale=0.48]{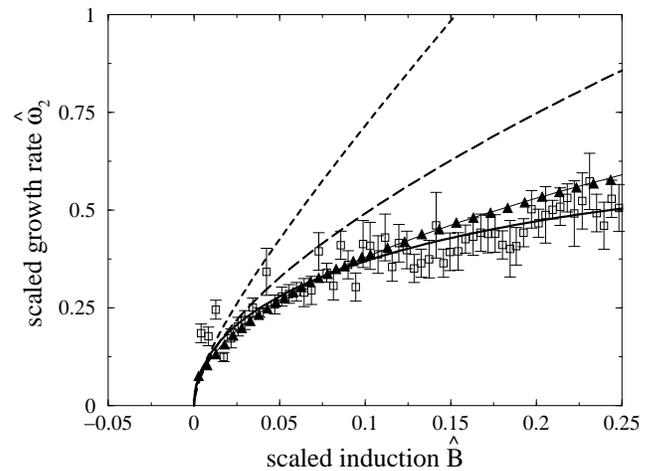}
  \caption{Scaled growth rate $\hat\omega_{2}$ versus the scaled
  induction $\hat B$ for the magnetic fluid EMG 909. The open squares
  give the experimental values with the corresponding errors. A fit
  for those data using the approximation Eq.~(3.9b)
  yields the thick solid line. Using a linear law of magnetization and
  an infinite thickness of the layer, the dashed line shows the
  theoretical result. The results with a nonlinear law of magnetization
  and a finite thickness of $h=5$ mm are indicated by the long-dashed
  line. From the numerical simulations the resulting growth rate is
  given by the filled triangles. A fit to these results with Eq.~(3.9b)
  gives the thin solid line.}
  \label{fig:growthrateEMG909_resultsPRE}
\end{center}
\end{figure}

\begin{figure}[tbph]
\begin{center}
  \includegraphics[scale=0.48]{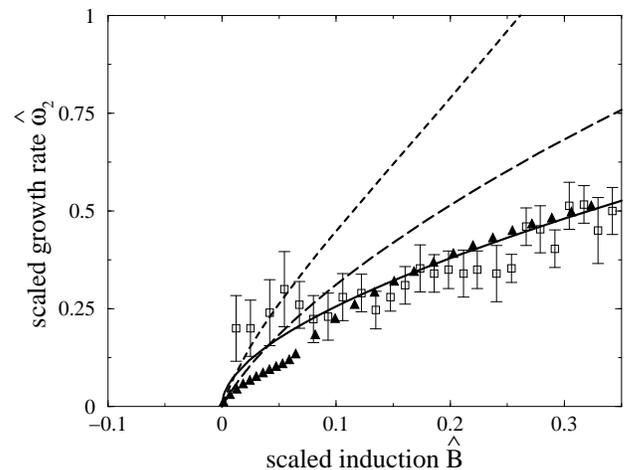}
  \caption{Scaled growth rate $\hat\omega_{2}$ versus
  the scaled induction $\hat B$ for the magnetic fluid APG J12.
  The symbols and types of lines are as in
  Fig.~\ref{fig:growthrateEMG909_resultsPRE}.}
  \label{fig:growthrateAPGJ12_resultsPRE}
\end{center}
\end{figure}

The measured growth rate is multiplied by the capillary time $t_c$
yielding the dimensionless variable $\hat\omega_{2}$. The
experimental values for EMG 909 (APG J12) are plotted as open
squares in Fig.~\ref{fig:growthrateEMG909_resultsPRE}
(Fig.~\ref{fig:growthrateAPGJ12_resultsPRE}). The size of the error
bars is mainly based on the uncertainty of the proper fitting range.
The four lines are results of theoretical considerations which will
be described in detail in the following sections.

\section{\label{sec:theory} Comparison with linear theory}

\subsection{\label{subsec:system} System and basic equations}

A horizontally unbounded layer of an incompressible, nonconducting,
and viscous magnetic fluid of thickness $h$ and constant density
$\rho$ is considered. The fluid is bounded from below by the bottom
of a container made of a magnetically impermeable material and has a
free surface with air above.

In a linear stability analysis, all small disturbances from the
basic state are decomposed into normal modes, i.e., into components
of the form $\exp [-i(\omega\,t -\vec q\,\vec r\,)]$, where $\vec r
=(x,y)$ and the wave number is the absolute value of the wave
vector, $q=|\vec q\,|$. With $\omega=\omega_1 +i\omega_2$, the real
part of $-i\omega$, $\omega_2$, is called the growth rate and
defines whether the disturbances will grow ($\omega_2>0$) or decay
($\omega_2<0$). The absolute value of the imaginary part of
$-i\omega$, $|\omega_1|$, gives the angular frequency of the
oscillations if it is different from zero. With the assumption that
the magnetization $\vec M$ of the magnetic fluid depends linearly on
the applied magnetic field $\vec H$,
$\vec M =(\mu_r -1)\vec{H}=\chi\vec{H}$, the linear stability analysis
leads to the dispersion relation \cite{weilepp96,abou97,mueller98}
\begin{eqnarray}
  \nonumber
  0 & = &\frac{\nu^2}{\tilde q\coth(\tilde q h)- q\coth (qh)}\biggr(
    \tilde q\left[ 4q^4 +(q^2+\tilde q^2)^2\right]\\
  \nonumber
  &&\times\coth(\tilde q h)-q\bigr[ 4q^2\tilde q^2
    +(q^2 +\tilde q^2)^2\bigr]\tanh (qh)\\
  \nonumber
  &&-\frac{4q^2\tilde q (q^2+\tilde q^2)}
    {\cosh (qh) \sinh (\tilde q h)}\biggr)
    +\tanh (qh) \biggr( g_0 q +\frac{\sigma}{\rho}\,q^3\\
  &&
    -\frac{\mu_0 \mu_r M^2}{\rho}
   \Lambda (qh)\,q^2\biggr)\, ,
  \label{eq:dispersion_relation}
\end{eqnarray}
where $\mu_r$ is the relative permeability of the MF, $M$ the absolute
value of the magnetization, $\vec g_0=(0,0,-g_0)$ the acceleration
due to gravity, $\mu_0$ the permeability of free space, $\tilde q
=\sqrt{q^2 -i\omega/\nu}$, and
\begin{equation}
\Lambda (qh) = \frac{{\rm e}^{qh} (1+\mu_r) + {\rm
e}^{-qh}(1-\mu_r)}{{\rm e}^{qh}(1+\mu_r)^2 - {\rm
e}^{-qh}(1-\mu_r)^2}.
\end{equation}
A nonlinear law of magnetization for a more realistic comparison
with the experiment is examined, too. The magnetic part of the
dispersion relation~(\ref{eq:dispersion_relation}) changes to
\begin{eqnarray}
\nonumber &&(1+\chi)M^2\Lambda (qh) \rightarrow (1+\bar\chi)M^2
\\
\label{eq:new_magnetic_part} && \hskip -1.3cm \times \left( \frac{{\rm
e}^{qh(1+\bar\chi)/(1+\chi)}(2+\bar\chi) -\bar\chi{\rm
e}^{-qh(1+\bar\chi)/(1+\chi)}}{{\rm
e}^{qh(1+\bar\chi)/(1+\chi)}(2+\bar\chi)^2 -\bar\chi^2{\rm
e}^{-qh(1+\bar\chi)/(1+\chi)} }\right)
\end{eqnarray}
with the differential susceptibility $\chi_d=(\partial M/\partial
H)_{H_g}$, the chord susceptibility $\chi_c=(M/H)_{H_g}$, and
$1+\bar\chi =\sqrt{(1+\chi_d)(1+\chi_c)}$ at a given strength of the
magnetic field $H_g$. With the help of the magnetization curve (see
Fig.~\ref{fig:magnetization}) one can determine $\chi_d$, $\chi_c$,
and $\bar\chi$ for every supercritical induction.

The condition of marginal stability, $\omega=0$, defines the critical
quantities at which the Rosensweig instability occurs. In the limit
of an infinitely thick ($h\rightarrow\infty$) layer, the critical
induction and the wave number, respectively, are
\begin{equation}
  \label{eq:critical_values}
  B_{c,\infty}^2 = \frac{2\mu_0\,\mu_r (\mu_r +1)\sqrt{\rho\,\sigma\,g}}{(\mu_r -1)^2},
  q_c =\sqrt{\frac{\rho\, g}{\sigma}}\, .
\end{equation}
These critical values for the {\it onset} of the instability apply
for viscous as well as for inviscid magnetic fluids.

\subsection{\label{subsec:growthrate}Growth rate of the most unstable
linear pattern for a linear law of magnetization}

Within the band of unstable wave numbers, the mode with the largest
growth rate is of primary importance. For its estimation it is
advantageous to consider the dimensionless form (indicated by the
bar) of the dispersion relation~(\ref{eq:dispersion_relation}) in
the limit $h\rightarrow\infty$ for growing disturbances, i.e.
$\omega= i\omega_2$ with $\omega_2>0$,
\begin{equation}
  \left(\bar\nu + \frac{\bar\omega_2}{2\bar q\, ^2}\right)^2
  +\frac{\bar q +\bar q\, ^3 -2 \bar B^2 \bar q\, ^2}{4 \bar q\, ^4}\\
  \label{eq:disprel_dimless}
  -\bar\nu^2\sqrt{ 1 + \frac{\bar\omega_2}{\bar\nu\bar q\, ^2}} =0\; .
\end{equation}
All lengths were scaled with $[\sigma/(\rho\, g_0)]^{1/2}$, the time
with $\sigma^{1/4}/(g_0^{3/4}\rho^{1/4})$, the viscosity with
$\sigma^{3/4}/(g_0^{1/4} \rho^{3/4})$, and the induction with
$B_{c,\infty}$. The maximal growth rate is determined by
$\partial\bar\omega_2/\partial\bar q =0$.

An expansion of $\bar B$, $\bar q$, and $\bar \omega_2$ in the form
\begin{equation}
  \label{eq:expansion_bar}
  \bar B = 1+\hat B,\;\;\;\;
  \bar q = 1+\hat q_m,\;\;\;\;
  \bar \omega_2 = 0 + \hat\omega_{2,m}
\end{equation}
leads to an analytical expression of the dependence of
$\hat\omega_{2,m}$ on the induction and the viscosity. All careted
quantities in (\ref{eq:expansion_bar}) are small $(\hat B, \hat
q_m, \hat\omega_{2,m}\ll 1$), and denote the scaled distances from
the critical values at the onset of the instability. If
$\bar\nu \gg \hat\omega_{2,m}$, Eq.~(\ref{eq:disprel_dimless})
and its derivative are expanded by means of higher-order terms of
the applied induction in the ansatz
\begin{eqnarray}
  \label{eq:ansatz_omega}
  \hat\omega_{2,m} &=& \alpha\hat B+\beta\hat B^2+\gamma\hat B^3 +O(\hat B^4)\; ,\\
  \label{eq:ansatz_q}
  \hat q_m &=& \delta\hat B^2 +\epsilon\hat B^3 +O(\hat B^4)\; .
\end{eqnarray}
The dependence of the maximal growth rate on the parameters viscosity
and induction is then given by \cite{lange01_growth}
\begin{equation*}
  \hat\omega_{2,m} =
  \begin{cases}
      \displaystyle{\frac{2}{\bar\nu}\hat B +\left( \frac{1}{\bar\nu}
      -\frac{3}{\bar\nu^3}\right)\hat B^2 + \left(\frac{10}{\bar\nu^5}
      -\frac{3}{\bar\nu^3}\right)\hat B^3 } \\[2ex]
      \hskip 2.5 cm
      \text{for } 0\leq\hat B< \bar\nu^2/6
      \hskip 0.9 cm
      \text{(3.9a)}\\[1ex]
      c_1\sqrt{\hat B} + c_2\hat B
      \hskip 0.5 cm
      \text{for } \bar\nu^2/6\ll\hat B\leq 0.4\; .
      \hskip 0.4 cm \text{(3.9b)}
  \end{cases}
\end{equation*}
For scaled inductions larger than $\bar\nu^2/6$, one has to solve the
full implicit dispersion relation~(\ref{eq:dispersion_relation}) and
its derivative with respect to $q$ numerically. The fit for an excellent
agreement with these numerical data includes a linear term and a
square-root term with respect to $\hat B$, where the coefficients
depend on the magnetic fluid.

\begin{figure}[htbp]
  \begin{center}
  \includegraphics[scale=0.45,angle=0]{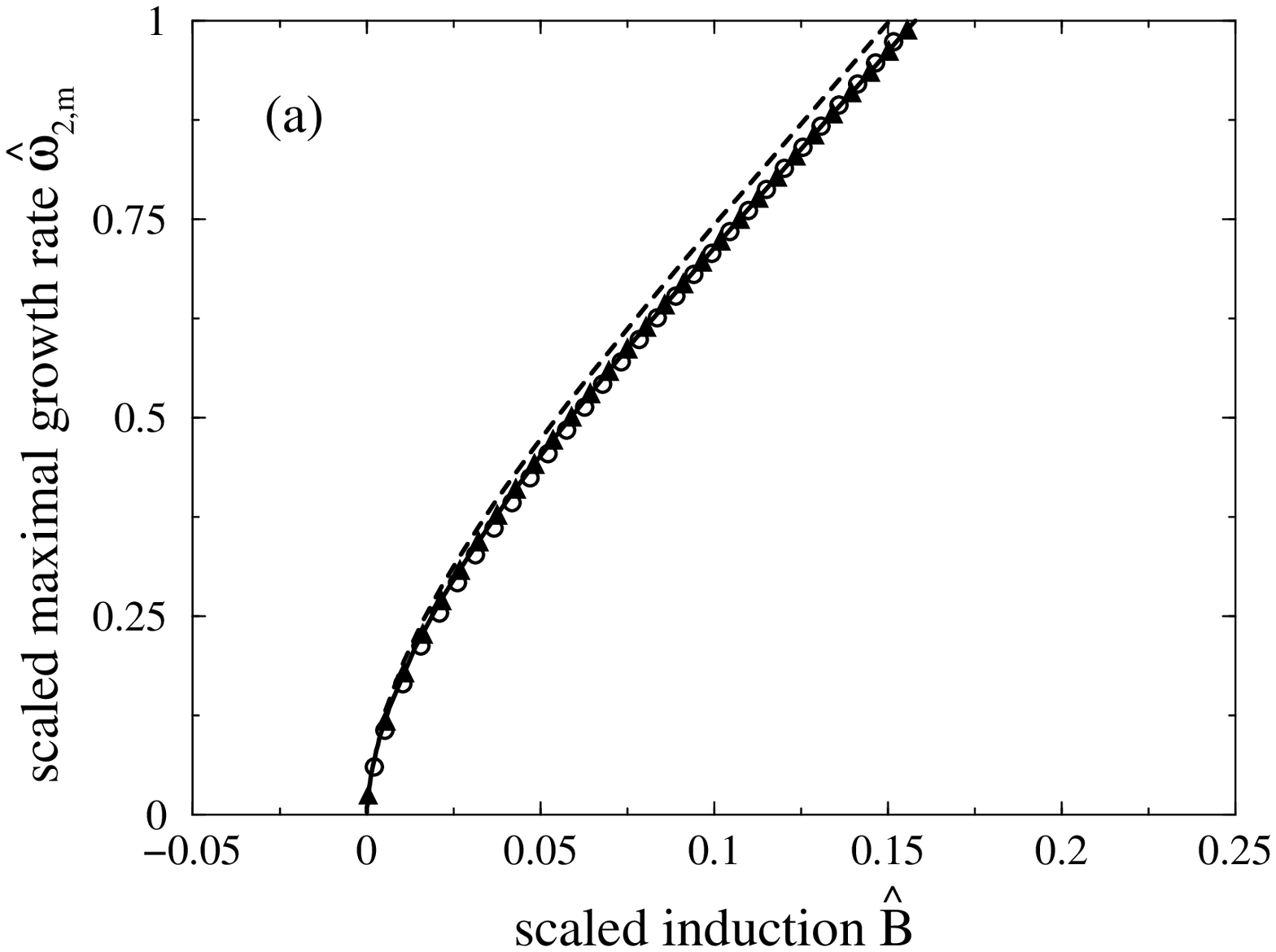}
  \phantom{i$\,$}
  \includegraphics[scale=0.45,angle=0]{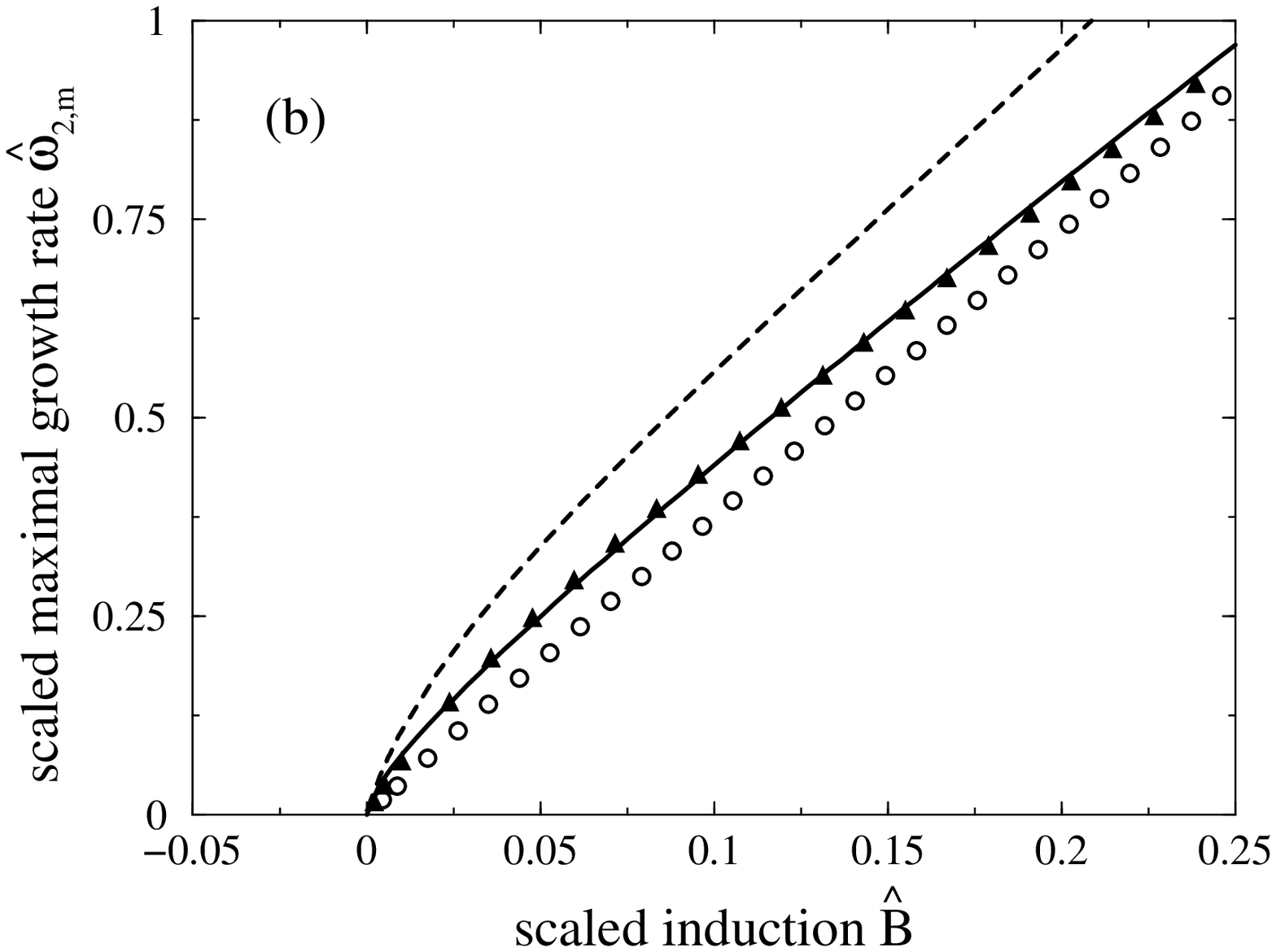}
  \caption{Scaled maximal growth rate $\hat\omega_{2,m}$ versus
  the scaled induction $\hat B$ for the magnetic fluids EMG 909 (a)
  and APG J12 (b). Using an infinite thickness of the layer, the
  solid lines shows the theoretical result for EMG 909 (APG J12).
  The results for a finite thickness of $h=5$ $(2.5)\,\mathrm{mm}$
  are indicated by filled triangles (open circles), respectively.
  A calculation with $h=5\,\mathrm{mm}$ and a dynamical viscosity
  reduced by 50\,\% gives the dashed lines.}
  \label{fig:growthrate_theo_linMag_various}
  \end{center}
\end{figure}

The calculation of the scaled induction, which separates the two scaling regimes in Eq.~(3.9),
gives ${\bar\nu}^2/6\simeq 8\times 10^{-5} (\simeq 9\times 10^{-3})$ for the fluid EMG 909 (APG
J12). Therefore Eq.~(3.9b) has to be used for most practical experiments because such supercritical
inductions above $B_c$ can hardly be accomplished in an experiment. Using the test fluids EMG 909
and APG J12, respectively, the fit of the maximal growth rate results in the coefficients
$c_1\simeq 1.39$ and $c_2\simeq 2.77$ for EMG 909 \cite{comment1_PRE06} and $c_1\simeq 0.45$ and
$c_2\simeq 2.97$ for APG J12. The corresponding curves are plotted as solid lines in
Figs.~\ref{fig:growthrate_theo_linMag_various}(a) and \ref{fig:growthrate_theo_linMag_various}(b),
respectively.

Next we test the robustness of the theoretical curve
against variations of the experimental parameters. Taking into
account the finite thickness of the layer does not create much
difference if the test fluid is EMG 909: neither a thickness of
$h=5$ mm (filled triangles) nor of $h=2.5$ mm (open circles) causes
much change, as shown in
Fig.~\ref{fig:growthrate_theo_linMag_various}(a). Figure
~\ref{fig:growthrate_theo_linMag_various}(b) displays that for the
fluid APG J12 only the smallest tested thickness of $2.5$ mm results
in an apparent difference in comparison to the case of an infinite
thickness. Additionally to the experimental filling level of $5$ mm,
the height of $2.5$ mm had been chosen because the inevitable field
gradient at the edge of the vessel can diminish the fluid level in
the central part by up to a factor of 0.6 \cite{reimann03}.

During the course of the experiment an increase of the temperature
of the MF may occur. Therefore a hypothetical reduction of the dynamic
viscosity $\eta$ by 50\,\% at a filling level of $h=5$ mm is considered
in order to test its influence. The results are indicated by the dashed
lines in Fig.~\ref{fig:growthrate_theo_linMag_various} and show a
noticeable influence on the maximal growth rate only in the case of
the fluid APG J12. All in all, the theoretical behavior seems to be rather
robust to variations of the experimental parameters.

\section{\label{sec:comparison_theory_exp}Comparison of
theoretical and experimental results}

The comparison starts with the values for EMG 909. The measured growth rates (see
Fig.~\ref{fig:growthrateEMG909_resultsPRE}, open squares) can be fitted using the
approximation~(3.9b) which is marked by the thick solid line in
Fig.~\ref{fig:growthrateEMG909_resultsPRE}. It results in the coefficients $c_{\rm 1,exp}\simeq
1.44$ and $c_{\rm 2,exp}\simeq -0.87$. The dashed line shows the result for a linear law of
magnetization, i.e. the numerical solution of the dispersion
relation~({\ref{eq:dispersion_relation}) and its derivative with respect to $q$, and an infinite
thickness of the layer. Applying again a fit according to Eq.~(3.9b) yields $c_{\rm
1,theo,lin}\simeq 1.39$ and $c_{\rm 2,theo,lin}\simeq 2.77$. Comparing these two curves and the
corresponding fit coefficients (see also Table~\ref{tab:table_B_and_c}), it becomes clear that
these theoretical values differ grossly from the measured ones.

\begin{table}[htbp]
\caption{List of critical inductions $B_c$ and fit coefficients $c_1$ and $c_2$
for EMG 909 and APG J12. The theoretical, numerical, and experimental data were
fitted according to Eq.~(3.9b), where $c_1$ scales the square-root term and $c_2$
the linear term.}
\begin{ruledtabular}
\begin{tabular}{lccc}
& $B_c$ (mT) & $c_1$ & $c_2$\\
\hline
\multicolumn{4}{c}{EMG 909} \\
Experiment                        & 25.7  & 1.44   & -0.87 \\
Theory, $M(H)$\,linear\footnotemark[1]   & 20.1  & 1.39   &  2.77 \\
Theory, $M(H)$\,nonlinear\footnotemark[2] & 24.9  & 1.24   &  0.94 \\
Numerics                          & 25.0  & 1.23   & -0.10 \\
\multicolumn{4}{c}{APG J12} \\
Experiment                        & 21.7  & 0.69   & 0.32 \\
Theory, $M(H)$\,linear\footnotemark[1]    & 17.3  & 0.45   & 2.97 \\
Theory, $M(H)$\,nonlinear\footnotemark[2] & 20.4  & 0.47   & 1.45 \\
Numerics                          & 21.9  & --     & --   \\
\end{tabular}
\end{ruledtabular}
\footnotetext[1]{The linear stability theory uses a linear function
to fit the magnetization.}
\footnotetext[2]{The linear stability theory uses the Langevin
function to fit the magnetization.}
\label{tab:table_B_and_c}
\end{table}

In Sec.~\ref{subsec:growthrate} we saw that taking into account a
finite layer thickness or a variation of the viscosity of the MF has
only a diminutive influence, and therefore cannot much reduce the
difference with the experimental data. Thus a nonlinear law of
magnetization is examined for a more appropriate comparison. Using
the actually measured material data, a finite layer thickness of
$h=5$ mm, and the magnetization curve of Fig.~\ref{fig:magnetization}(a)
results in the data plotted by the long-dashed line. The latter,
which can be fitted by Eq.~(3.9b) using
$c_{\rm 1,theo,nlin}\simeq 1.24$ and $c_{\rm 2,theo,nlin}\simeq 0.94$,
lies appreciably closer to the experimental data.

Figure\,\ref{fig:growthrateAPGJ12_resultsPRE} shows the experimental
(open squares) and the theoretical results for the second tested
magnetic fluid, APG J12. A fit of the experimental data by means of
Eq.~(3.9b) gives the thick solid line, where the
fit coefficients are given in Table~\ref{tab:table_B_and_c}. The
theoretically determined growth rates are based on either a linear law
for the magnetization (dashed line) or a nonlinear one (long-dashed line).

In contrast to previous studies \cite{lange00_wave,reimann03}, a
nonlinear law of magnetization is necessary in order to reduce the
gap between the theoretical data for the maximal growth rate and the
experimental results. Despite that, for $\hat B=0.25$ the theoretical
value thus estimated is about 70\,\% (EMG 909) and 35\,\% (APG J12),
respectively, above the measured one.

One may discuss several reasons for the considerable disagreement between
theory and experiment, such as errors in the material parameters or a
limited resolution of the sensor array. However, most importantly a
systematic deviation may have its origin in the finite size of the
container: because of that, experiment and theory may have different
starting conditions. Figure~\ref{fig:finite_size} displays a
radioscopic surface profile recorded for $\hat B=-0.1$, i.e. in a
subcritical region of the instability. One clearly sees surface
undulations well before the critical induction. They are most
prominent next to the edge of the vessel and have their origin in
the discontinuity of the magnetization at this place. Thus, the
experiment will start with a finite disturbance whereas the theory
is estimated for a infinitesimal perturbation.

\begin{figure}[htbp]
  \begin{center}
  \includegraphics[width=0.95\columnwidth,angle=0]{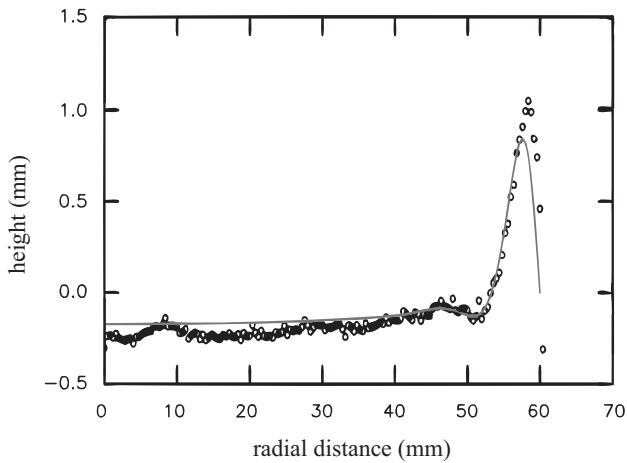}
  \caption{Radioscopic measured surface profile (circles) of the
  fluid EMG 909 recorded for $\hat B=-0.1$ for a fluid height of 3 mm.
  At the position of 0 mm is the center, at 60 mm the inner edge
  of the Teflon vessel. The \emph{y} axis denotes the height of the fluid
  with respect to its level without a magnetic field.}
  \label{fig:finite_size}
  \end{center}
\end{figure}

In the following we perform numerical calculations startingwith a
finite perturbation, in order to test whether this can better
describe the experimental data.

\section{\label{sec:numerics}Numerical simulations}

Our numerical simulations are based on a coupled system of
nonlinear governing equations: the Maxwell equations in the magnetic
liquid and its surroundings, the Navier-Stokes equations in the
magnetic liquid, and the Young-Laplace equation on the free surface.

Because magnetic fluids can be regarded as insulators, the Maxwell
equations in the entire space are given by
\begin{equation}
\label{MaxwellReal}
\text{curl}\,\vec H = \vec 0,\quad
\text{div}\,\vec B = 0 \, ,
\end{equation}
with the constitutive relation
\[
\vec B = \begin{cases}
\;\mu_0 (\vec M + \vec H) & \text{in }\Omega_F(t),\\[1ex]
\;\mu_0 \vec H & \text{outside }\Omega_F(t),
\end{cases}
\]
where $\Omega_F(t)$ denotes the domain that is occupied by the
magnetic liquid at time $t$. The magnetization $\vec M$ is assumed
to follow a Langevin law [see Eq.~(\ref{eq:langevin})]. Such a
nonlinear law results in a better approximation of the measured
magnetization than a linear dependence of $\vec M$ on $\vec H$ as
used in Sec. \ref{sec:theory}.

The hydrodynamic behavior of the magnetic liquid is described by the
nonstationary, incompressible Navier-Stokes equations in the
time-dependent fluid domain $\Omega_F(t)$. These equations read as
follows:
\begin{subequations}
\label{NSE}
\begin{align}
\rho\left(\frac{\partial \vec u}{\partial t}
+ (\vec u\cdot\nabla)\vec u\right)
&  = \text{div }\mathbb{T}(\vec u,p,\vec H) - \rho \vec g_0 z \, ,
\label{NSE1}\\
\text{div }\vec u & = 0 \, .
\label{NSE2}
\end{align}
\end{subequations}
Here, $\vec u$ denotes the
fluid velocity, $p$ the sum of the hydrodynamic pressure and the
fluid-magnetic pressure, and $\mathbb{T}$ the magnetically augmented
stress tensor with
\begin{align*}
\mathbb{T}_{ij}(\vec u,p,\vec H) = & \eta\left(
\frac{\partial u_i}{\partial x_j} + \frac{\partial u_j}{\partial x_i}
\right)
- \left( p + \frac{\mu_0}{2} H^2 \right)\delta_{ij}\\
& + B_i H_j \, .
\end{align*}
The system of equations is completed by the force balance at the
free surface which is given by the Young-Laplace equation in the
following form
\begin{equation}
\label{YL}
[\mathbb{T}(\vec u,p,\vec H)\vec n] = \sigma\mathcal{K}\vec n \, ,
\end{equation}
where $\sigma$ is the surface tension, $\vec n$ the outer unit normal
on $\partial\Omega_F(t)$, and $\mathcal{K}$ the sum of the principal
curvatures. Here, $[\psi]$ denotes the jump of the quantity $\psi$
across the interface. Furthermore, the kinematic condition
\begin{equation}
\label{kincond}
\vec u\cdot\vec n = v_\Gamma
\end{equation}
with the normal velocity $v_\Gamma$ of the free surface $\Gamma_F$
is used. Finally, the system is closed with initial and boundary
conditions.

In order to solve the coupled system of nonlinear partial differential
equations numerically, it is split into two subproblems: a magnetostatic
problem for the magnetic field and a flow problem which also involves
the Young-Laplace equation.

We consider for our numerical simulations a bounded three-dimensional
domain $\widetilde{\Omega}=\widetilde{G}\times(\widetilde{z}_b,
\widetilde{z}_t)$ with a two-dimensional hexagonal base $\widetilde{G}$
which contains exactly one peak. Furthermore, the interval
$(\widetilde{z}_b,\widetilde{z}_t)$ in the $\widetilde{z}$ direction is
chosen such that its end points are far below and above the free surface,
respectively. This ensures that the position of the
free surface does not affect the magnetic field on the upper and lower
boundaries.

The Maxwell equations are transformed into their dimensionless form by
using the strength of the applied magnetic field and a characteristic
length scale $l$, which is a fixed multiple of the wavelength of the
pattern. In this way, the domain $\Omega=G\times(z_b,z_t)$ is obtained.
The Maxwell equations in dimensionless form read
\begin{equation}
\label{Maxwell}
\text{curl }\vec{H} = \vec{0},\quad \text{div }\vec{B}=0
\qquad\text{in }\Omega \, .
\end{equation}
The first differential equation in~\eqref{Maxwell} ensures the
existence of a scalar magnetostatic potential $\varphi$ such that
$\vec{H} = -\vec{\nabla}\varphi$. Hence, by using the second differential
equation of~\eqref{Maxwell}, we get
\begin{equation}
\label{magpot}
-\text{div}\big[\mu(\vec{x},|\nabla\varphi|)\nabla\varphi\big] = 0
\qquad\text{in }\Omega \, .
\end{equation}
The coefficient function $\mu(\vec{x},H)$ is given by
\[
\mu(\vec{x},H) = \begin{cases}
1 & \vec{x}\in\Omega_A(t),\\[1ex]
\displaystyle 1 + \frac{M(H)}{H} & \vec{x}\in\Omega_F(t),
\end{cases}
\]
where $\Omega_F(t)$ and $\Omega_A(t)$ are the three-dimensional subdomains
of $\Omega$ that correspond to the areas inside and outside the magnetic
liquid at time $t$, respectively. Eq.~\eqref{magpot} is equipped with
boundary conditions which correspond to the case of a flat surface. We
refer to~\cite{gollwitzer07} for details.

The solution of the magnetostatic problem~\eqref{magpot} is
approximated by a finite-element method with continuous, piecewise
triquadratic functions. The nonlinearity in~\eqref{magpot} due
to the nonlinear magnetization law is overcome by a fixed-point
iteration. In each iteration step, the large system of linear
equations arising is solved by a geometric multigrid method.

For solving the time-dependent Navier-Stokes equations, we start with a
semidiscretization in time by applying the fractional-step
$\vartheta$-scheme~\cite{BGP87,Glo03}, which is of second order and
strongly A-stable~\cite{KR94,MU93}. The resulting equations in each
time step are solved by a finite-element method which incorporates the
Young-Laplace equation~\eqref{YL}. Furthermore, the arbitrary Lagrangian
Eulerian (ALE) approach is applied to handle the time-dependent fluid
domains.

It is well known that the finite-element spaces which are used to
approximate velocity and pressure in the discretized Navier-Stokes
equations cannot be chosen independently but have to satisfy a
constraint that is given by the inf-sup (or Babu\v{s}ka-Brezzi)
condition. We used in our calculations continuous, piecewise triquadratic
functions for the velocity and discontinuous, piecewise linear functions
of the pressure. This pair of spaces satisfies the inf-sup
condition~\cite{GR86,MT02}.

After discretizing the Navier-Stokes equations in time and space,
one has to solve in each time step a nonlinear saddle-point problem.
The nonlinearity is resolved by a fixed-point iteration. The resulting
system of linear equations is again solved by a geometric multigrid method.
We refer to Refs.\,\cite{JM01,Joh02,Joh04} for details.

The position of the free surface is updated after each time step
by using the kinematic condition~\eqref{kincond}. Since the domain
that is occupied by the magnetic liquid changes in time, the meshes
used by both finite-element methods have also to change in time in
order to guarantee that the free surface is approximated by faces of
three-dimensional mesh cells. We have used a simple algebraic mesh
update which arranges the mesh points according to the height of the
free surface position.

All numerical results were obtained by using the software package
MooNMD~\cite{JM04}.

In order to get the developed surface profile, one has to choose a
proper initial surface perturbation. Starting with a completely flat
surface ($z\equiv 0$), the calculations will result in the same flat
surface for all times, independent of the strength of the applied
magnetic field. We used a rotationally symmetric cosinelike profile
as initial perturbation. Its amplitude was selected as 0.007 mm
(0.034 mm) for the fluid EMG 909 (APG J12), respectively.
\begin{figure}[htbp]
  \begin{center}
  \includegraphics[width=\columnwidth,angle=0]{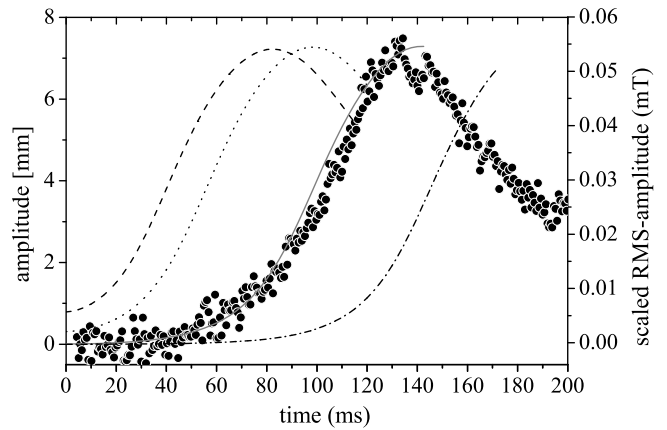}
  \caption{Temporal evolution of the measured peak amplitude for
  $\hat B$ = 0.25 for the MF EMG 909 (dots) and the corresponding
  evolution of the calculated peak amplitude for different initial
  perturbation heights of 0.791 (dashed line), 0.313 (dotted line),
  0.007 (full line), and 0.001 mm (dashed-dotted line). The scaled
  rms amplitude is the measured rms amplitude minus its offset at 0 ms.}
  \label{fig:EMG909_starting_amplitude}
  \end{center}
\end{figure}
Figure~\ref{fig:EMG909_starting_amplitude} demonstrates for the fluid
EMG 909 that higher (lower) starting values result in an earlier (later)
growth of the perturbation in comparison with the experimental curve.
The selected perturbation gives the expected dynamic growth of the
perturbation into the oscillatory relaxation process, provided the
strength of the applied field is large enough.
Note that the obtained dynamic growth rate is independent of the
initial perturbation height.

It has been shown in theory and experiment that the wave number of
maximal growth depends linearly on the scaled magnetic induction
$\hat B$ \cite{lange00_wave}. For a first attempt to unravel the
mismatch between theory and experiment, we performed all numerical
calculations with the critical wavelength $q_c$.

From these numerical simulations, we obtain a critical value for the
onset of the Rosensweig instability by taking the smallest value that
results in a growth of the perturbation. If the strength of the applied
field is smaller than this obtained threshold, then the initial surface
perturbation declines towards a flat surface. The sets of critical
inductions for the two fluids are collected in the second row of
Table~\ref{tab:table_B_and_c}.

Also from numerical simulations it is possible to determine the growth
rate. Due to the lack of noise, the fitting range for the exponential
growth of the amplitude can be easily determined via the maximum of the
numerical differentiated amplitude curve. The resulting values of the
growth rate at different supercritical inductions are indicated by
filled triangles in Fig.~\ref{fig:growthrateEMG909_resultsPRE} and
Fig.~\ref{fig:growthrateAPGJ12_resultsPRE}. Fitting these numerical
results for the fluid EMG 909 with Eq.~(3.9b)
results in the coefficients $c_{\rm 1,num}\simeq 1.23$ and
$c_{\rm 2,num}\simeq -0.1$ (see the thin solid line). Due to the structure
of the numerical results for the fluid APG J12, we refrained from a
single fit over the entire range of $\hat B$. Therefore no fit
coefficients $c_{\rm 1,num}$ and $c_{\rm 2,num}$ for APG J12 are
given in the corresponding list (Table~\ref{tab:table_B_and_c}).

\section{\label{sec:discussion}Discussion and Conclusions}

We performed measurements of the growth of surface undulations at
the Rosensweig instability for different supercritical inductions,
applied to two magnetic fluids of different viscosity. Comparing
the values of the growth rates for both tested magnetic fluids
(cf. Table\,\ref{tab:table_B_and_c}), one notes that the less
viscous one (EMG 909) has larger growth rates than the more viscous
one (APG J12). At $\hat B=0.25$ the experimental value of
$\hat\omega_{\rm 2,m}$ for EMG 909 is about $18$\% larger than the
corresponding value for APG J12. That the less viscous fluid grows
faster is intuitively clear since less viscosity goes along with
less friction inside the fluid. Therefore more energy is transformed
into the movement of the fluid, which appears in our case as the
growth of the peaks.

A comparison of experimental and theoretical values (cf.
Figs.~\ref{fig:growthrateEMG909_resultsPRE} and
\ref{fig:growthrateAPGJ12_resultsPRE}) shows that the theoretical
values, obtained from calculations with a linear magnetization curve,
overestimate the experimental ones considerably. This mismatch
could be reduced by taking into account the proper nonlinear
magnetization curve in the linear theory. Even so, the estimated
growth rates remained 70\,\% (35\,\%) above the experimental values
for the less (more) viscous fluid, respectively. There are several
reasons for this discrepancy.

First, we do not measure the growth of only the fastest-growing mode,
but an averaged growth of several modes, by using the rms value of
the measured amplitude from the Hall-sensor array. In contrast to the
case of the static, tilted field instability \cite{reimann05}, we
could not fit the spatial modulation of the signal of the sensor array
with periodic functions. This difficulty might stem from the higher
complexity of the evolving pattern which can not fully be captured by
a one-dimensional array, and the limited spatial resolution of the array.
As an outcome we are not able to estimate a dispersion relation
$\omega_2(q)$, as in Refs.\,\cite{voeltz01,schroeter02}. Therefore
the growth rate extracted from the rms values of the magnetic amplitude
data can only be considered a rough estimate for a maximal growth rate
determined from the dispersion relation.

Second, the vessel in the experiment has a finite size, which causes
an inhomogeneous growth of the surface amplitude starting with a
\textit{finite amplitude} from the edge of the vessel. In contrast,
the theory is for a laterally infinite layer of fluid and infinitesimal
surface perturbations. We could corroborate this thesis with radioscopic
measurements of the static surface profile, unveiling a finite surface
elevation for subcritical inductions.

Here the numerical simulation via the finite amplitude method comes
to the rescue, because it can take a starting condition with finite
amplitude into account. The calculated temporal evolution of the
surface undulations agrees well with the measurement, including the
oscillations, which were observed for two different viscosities.
This feature is beyond the framework of a linear stability analysis
and can be calculated only with the help of numerical methods
\cite{matthies05}. More importantly, the numerically estimated growth
rates match the measured ones well. For supercritical inductions of
$\hat B\leq 0.1$, the agreement between experimental and numerical
values is clearly better for the less viscous fluid EMG 909. The two
data sets can hardly be distinguished. For supercritical inductions
of $\hat B> 0.1$, the agreement between experimental, numerical, and
theoretical values is clearly better for the more viscous fluid
APG J12. The numerical results fall practically onto the fit of the
experimental ones (compare filled triangles and thick solid line in
Fig.~\ref{fig:growthrateAPGJ12_resultsPRE}).

Remaining discrepancies between experiment and numerics may stem
from the following. Due to computational costs, so far the numerical
simulations were performed for a pattern with fixed, critical wavelength
$q_c$ for all values of the magnetic induction. In future, more
refined calculations will take into account the linear dependence
$\omega_2 (q)$ for the wave number of maximal growth. Furthermore, in
the experiment first circular ridges appear, which then arrange in a
hexagonal pattern during their growth. This might change the growth
rate, in contrast to the numerical evolution, which starts already
with a hexagonal pattern. This latter point is difficult to solve
numerically.

For future experiments the finite amplitude at the beginning of the
experiment should be reduced, e.g. by introducing a ramp as in
Ref.~\cite{gollwitzer07}. More importantly it will be necessary to
reduce, to the highest possible extent, the effect of the lateral
boundaries on the growth of the unstable mode by choosing improved
experimental and computational conditions (e.g. size of the container).
Moreover, we expect an improvement of the accuracy by a radioscopic
measurement of the growth rate with a two-dimensional x-ray detector
\cite{richter01}, becoming feasible for slow evolution of highly
viscous magnetic fluids. A Fourier analysis of these spatiotemporally
resolved surface profiles will allow an estimation of the growth rate
of the fastest-growing mode.

To conclude, we have experimentally, theoretically, and numerically
investigated the growth rate during the first stage of pattern
formation in the Rosensweig instability. Despite the use of a
nonlinear law of magnetization there remains a discrepancy between
the predictions of linear stability analysis and experimental data. In
contrast, the experimental data are confirmed by numerical
simulations using a nonlinear magnetization curve together with a
finite initial surface undulation. The growth behavior of the
related electrostatic instability should be similar, but remains to
be investigated.

\section*{Acknowledgements}

The authors would like to thank Achim Beetz for taking the photo in
Fig.~1, the Elektronik Workshop (ELUB) for developing the 32-channel
amplifier, Christian Gollwitzer for measuring the magnetization
curves, Bert Reimann for building the experimental setup, and
Konstantin Morozov and Lutz Tobiska for discussion. The work was
supported by the Deutsche Forschungsgemeinschaft under Grants
No. Ri 1054/1 and No. La 1182/2.

\end{document}